\documentclass[prb,showpacs,amssymb]{revtex4}

\usepackage{graphicx}

\begin{document}

\title{Decoupling Transition I. Flux Lattices
 in Pure Layered Superconductors \\}

\author{Ruth Goldin$^{1}$ and Baruch Horovitz$^{1,2}$}
\affiliation{$^1$Department of Physics and $^2$Ilze Katz center
for nanotechnology, Ben-Gurion University of the Negev, Beer-Sheva
84105, Israel}
\begin{abstract}
We study the decoupling transition of flux lattices in a layered
superconductors at which the Josephson coupling $J$ is
renormalized to zero. We identify the order parameter and related correlations; the latter are shown to decay as a power law in the decoupled phase.
Within 2nd order renormalization group we find that the transition is
always continuous, in contrast with
results of the self consistent harmonic approximation. The
critical temperature for weak $J$ is $\sim 1/B$, where $B$ is the
magnetic field, while for strong $J$ it is $\sim 1/\sqrt{B}$ and is
strongly enhanced. We show that renormaliztion group can be used to evaluate the Josephson plasma frequency and find that for weak $J$ it is $\sim 1/BT^2$ in the decoupled phase.
\end{abstract}


\maketitle
\newcommand{\rr}{\mbox{${\bf r}$}}
\newcommand{\RR}{\mbox{${\bf R}$}}
\newcommand{\QQ}{\mbox{${\bf Q}$}}
\newcommand{\uu}{\mbox{${\bf u}$}}
\newcommand{\qq}{\mbox{${\bf q}$}}


\section{Introduction}

 In many aspects the behavior of the high-temperature superconductors
  in the mixed state differs from that of conventional
superconductors. A combination of elevated critical temperatures,
high anisotropy and short superconducting coherence lengths
considerably enhance the role of  fluctuations on the vortex
interaction, which result in a noticeable change in the nature of
the mixed state.

The phase diagram of these layered superconductor in a magnetic
field $B$ perpendicular to the layers has been studied in
considerable detail \cite{kes}. A first order transition in
$YBa_{2}Cu_{3}O_{7}$ (YBCO) and in $Bi_{2}Sr_{2}CaCu_{2}O_{8}$
(BSCCO)  has been interpreted as  a flux lattice melting. The data
suggests that the first order line terminates at a multicritical
point, which for BSCCO is at $B_{0}\approx 300-10^{3}G$ and at
temperatures $T_{0}\approx 40-50 K$ \cite{khaykovich,fuchs}, while
for YBCO \cite{deligiannis} it is at $B_{0}\approx 2-10 T$ and
$T_0 \approx 60-80 K$, depending on disorder and Oxygen
concentration. At low temperatures as $B$ is increased a "second
peak" transition is manifested as a sharp increase in
magnetization. The second peak transition is at $B\approx B_{0}$
starting from the critical point; it is weakly temperature
dependent and shifts to lower fields with increasing disorder
\cite{khaykovich}. More recent data on BSCCO shows \cite{avraham},
however, that the second peak line connects smoothly with the
first order line; the presence of a multicritical point depends
then on possibility that an additional line joins in the high
field regime. Josephson plasma resonance studies
\cite{shibauchi,matsuda} have shown a significant reduction of the
Josephson coupling at the second peak transition. The combination
of reduced Josephson coupling and enhanced pinning indicates an
unusual phase transition.

The flux lattice can undergo a transition which is unique to a
layered superconductor, i.e. a decoupling transition
\cite{glazman1,glazman2,daemen,blatter}. In this transition the Josephson
coupling between layers vanishes while the lattice is maintained
by the magnetic coupling. The theory of Daemen et al.
\cite{daemen} employed the method of self consistent harmonic
approximation (SCHA) to find the decoupling temperature $T_d(B)$.
The SCHA leads to a conceptual difficulty since it predicts that
the average $\langle \cos \phi \rangle$, where $\phi$ is the Josephson phase, vanishes at $T>T_d$. Koshelev has shown \cite{koshelev} that $\langle \cos \phi \rangle$ is finite at all temperatures and in fact accounts for the experimentally observed Josephson plasma resonance at high temperatures. Thus the decoupling transition, as found by
SCHA, needs to be reinterpreted. The correct order parameter is in fact a non-local one (Eq. \ref{Q} below) and corresponds to the helicity modulus \cite{teitel} or to the critical current \cite{koshelev}.

In the present work we expand our earlier presentation \cite{h1}
and study the decoupling transition by 2nd order renormalization
group (RG). Section II defines the model and its effective Hamiltonian. In section III we identify the order parameter and related correlations and show that the latter decay either exponentially in the coupled phase or as a power law in the decoupled phase. We also show that RG can be used to evaluate $\langle \cos \phi \rangle$ and hence the Josephson plasma frequency and predict its behavior in the decoupled phase (section V).

Our RG analysis (section IV) shows that the decoupling transition is continuous
in the whole parameter range , in contrast with the SCHA result
that it is 1st order above some critical value of the Josephson
coupling \cite{daemen}. We trace the latter result to a deficiency
of the SCHA which does not allow for a scale dependent effective
mass (section VI).

Our analysis assumes that point defects like vacancies and
interstitials (V-I) of the flux lattice are not present. In
general, however, V-I defects are a relevant perturbation at
decoupling \cite{dodgson,ledou}. The effect is rather weak for
actual system parameters, as discussed in section VII.

The role of point disorder on the decoupling has been studied
separately \cite{h1,h2} and in more detail in a companion
article \cite{h4}. This study shows that disorder induced
decoupling leads to an apparent discontinuity in the tilt modulus
which then leads to an enhanced critical current and reduced
domain size. These results are in accord with the Josephson plasma
resonance \cite{shibauchi,matsuda} and other data on the second
peak transition. Of further interest is the effect of columnar
defects on the decoupling transition \cite{morozov}, an effect
which can further identify this transition.

\section{The model}

Consider a layered superconductor with a magnetic field
perpendicular to the layers. The flux lattice can be considered as
point vortices in each superconducting layer which are stacked one
on top of the other. Each point vortex, or a pancake vortex,
represents a singularity of the superconducting order parameter,
i.e. the order parameter's phase in a given layer changes by
$2\pi$ around the pancake vortex. These pancake vortices are
coupled by their magnetic field as well as by the Josephson
coupling between nearest layers.

The basic model for studying layered superconductors is the
Lawrence-Doniach \cite{lawrence} Hamiltonian in terms of
superconducting phases $\varphi_{n}({\bf r})$ on the $n$-th layer
and the vector potential ${\bf A}({\bf r},z),\, A_{z}({\bf r},z)$
(vectors such as ${\bf A}$ and ${\bf r}$ are 2-dimensional
parallel to the layers):
\begin{eqnarray}
{\cal H}^{LD}&=&\frac{1}{8\pi}\int d^{2}{\bf r} d z
 \left[
[{\mbox{\boldmath $\nabla$}}\times{\bf A}({\bf r},z) ]^{2}+
\frac{d}{\lambda_{a b}^{2}}\sum_{n}
(\frac{\Phi_{0}}{2\pi}{\mbox{\boldmath $\nabla$}}\varphi_{n}({\bf
r})-{\bf A}({\bf r},z))^{2} \delta(z-n d) \right]    \nonumber \\
              &-&{\tilde J}/\xi^{2}\sum_{n}\int d^{2}{\bf r}
\cos\left[          \varphi_{n}({\bf r})-\varphi_{n-1}({\bf r})-
\frac{2\pi}{\Phi_{0}} \int_{(n-1) d}^{nd} A_{z}({\bf r},z) d z
\right] \label{HLD}
\end{eqnarray}
where  $\lambda_{ab}$ and $\xi$ are the penetration length and
coherence length parallel to the layers, respectively, $d$ is the
spacing between layers and $\Phi_{0}=hc/2e$ is the flux quantum.
The first term in Eq.\ (\ref{HLD}) is the magnetic energy, the
second is the supercurrent energy while the last one is the
Josephson coupling where ${\tilde J}$ is the Josephson coupling
energy per area $\xi^2$ between neighboring layers. The model Eq.
(\ref{HLD}) qualifies as the standard model for layered
superconductors.

In this section we outline the derivation of an effective
Hamiltonian, which is the basis for RG analysis in the next
section. The partition sum for Eq.\ (\ref{HLD}) involves
integrating over both $\varphi_{n}({\bf r})$ and ${\bf A}({\bf
r},z)$, subject to a gauge condition.
 Since ${\bf A}({\bf r},z)$ is a Gaussian field (choosing the axial
gauge $A_{z}({\bf r},z)=0$) we can shift
 ${\bf A}\rightarrow {\bf A}+\delta {\bf A}$ where ${\bf A}({\bf
r},z)$ now satisfies the saddle point equations for its $x$, $y$
components
\begin{equation}
{\mbox{\boldmath $\nabla$}}\times{\mbox{\boldmath $\nabla$}}\times
{\bf A}({\bf r},z)=\frac{d}{\lambda_{a b}^{2}}\sum_{n} [
\frac{\Phi_{0}}{2\pi}{\mbox{\boldmath $\nabla$}}\varphi_{n}({\bf
r})-{\bf A}({\bf r},z) ]\delta(z-n d) \label{Maxwell}
\end{equation}
and then fluctuations in $\delta {\bf A}$ decouple from  those of
$\phi_{n}({\bf r})$. The partition sum at temperature $T$ involves
therefore integration only on the $\phi_{n}({\bf r})$ variables,
\begin{equation}
{\cal Z}= \int {\cal D}\phi_{n}({\bf r}) \exp [-{\cal H}^{LD}/T]
\end{equation}
with ${\bf A}({\bf r},z)$ in Eq.\ (\ref{HLD}) given by the
solution of Eq.\ (\ref{Maxwell}) for each configuration of
$\phi_{n}({\bf r})$ . Note that since Eq.\
(\ref{Maxwell}) is gauge invariant under ${\bf A}\rightarrow {\bf
A}-\frac{\Phi_{0}}{2\pi}{\bf\nabla}\chi({\bf r},n d)$ and
 $\varphi_{n}({\bf r}) \rightarrow  \varphi_{n}({\bf r})-\chi({\bf r},n d)$
 one can in  fact  choose any gauge.

We now decompose  $ \varphi_{n}({\bf r})$ to
\begin{eqnarray}
 \varphi_{n}({\bf r}) &=& \varphi_{n}^{0}({\bf r})+
\sum_{{\bf r} '}s_{n}({\bf r} ')\alpha({\bf r}-{\bf r} ')
\label{decompose}\\
\theta_{n}({\bf r})&=& \varphi_{n}^{0}({\bf r})-
\varphi_{n-1}^{0}({\bf r})
\end{eqnarray}
where $\varphi_{n}^{0}({\bf r})$ is the nonsingular part of
$\varphi_{n}({\bf r})$, $\alpha({\bf r})=\arctan (r_2/r_1)$ is the
angle at $\rr=(r_1,r_2)$, $s_{n}({\bf r})=1$ at pancake vortex
sites and $s_{n}({\bf r})=0$ otherwise. The sum in Eq.\
(\ref{decompose}) is then a sum on ${\bf r}'$ being the possible
vortex positions on the n-th layer, e.g. a grid with spacing
$\xi$.

Solving Eq.\ (\ref{Maxwell}) for ${\bf A}$ in terms of
$\theta_{n}({\bf r})$ and $s_{n}$, substituting in Eq.  (\ref{HLD})
yields after a straightforward analysis the equivalent Hamiltonian
\cite{h3}
\begin{equation}\label{H1}
{\cal H}^{(1)}={\cal H}_{v}+{\cal H}_{J}+{\cal H}_{f} \,,
\end{equation}
where  ${\cal H}_{v}$ is the vortex-vortex interaction via the 3D
magnetic field,  ${\cal H}_{J}$ is interlayer Josephson coupling
and ${\cal H}_{f}$ is an energy due to fluctuations of the
nonsingular  phase:
\begin{eqnarray}
  {\cal H}_{v}&=&\frac{1}{2}\sum_{r,n} \sum_{r',n'}s_{n}({\bf r})
G_{v}({\bf r}-{\bf r} ';n-n')s_{n'}({\bf r} ') \label{Fv}\\
{\cal H}_{J}&=&- \frac{{\tilde J}}{\xi^{2}}\sum_{n} \int d^{2}{\bf
r} (\cos[\theta_{n}({\bf r}) + \sum_{{\bf r} '}(s_{n}({\bf r}
')-s_{n-1}({\bf r} ')) \alpha({\bf r}-{\bf r} ')]-1)
\label{FJos}\\
    {\cal H}_{f}&=&\frac{1}{2}\int \frac{d^2qdk}{(2\pi)^3} G_{f}^{-1}(q,k)
|\theta({\bf q},k)|^{2}                            \label{Ff}
\end{eqnarray}
Here $({\bf q},k)$ is a 3D wave vector which is the Fourier
transform to ${\bf r}, z$ with $|q|<1/\xi$ and $|k|<\pi/d$, and
\begin{eqnarray}
G_{v}(q,k)&=&\frac{\Phi_{0}^{2}d^{2}}{4\pi\lambda_{a b}^{2}}
\frac{1}{q^{2}}\frac{1}{1+f(q,k)}          \\
    f(q,k)&=&\frac{d}{4\lambda_{a b}^{2}q}\frac{\sinh qd}
{\sinh^{2}\frac{qd}{2}+\sin^{2}\frac{kd}{2}}   \\
G_{f}(q,k)&=&\frac{16\pi^{3}d^{2}}{\Phi_{0}^{2}q^{2}}\left(1+
\frac{4\lambda_{a b}^{2}} {d^{2}} \sin^{2}\frac{kd}{2}\right)\,.
\end{eqnarray}

Consider now a flux lattice with equilibrium positions ${\bf
R}_{l}$ (e.g. a hexagonal lattice) where $l$ labels the flux
lines. The actual positions of the pancake vortices deviate from
the perfect lattice positions by ${\bf u}_{l}^{n}$ on the $n$-th
layer. The function $s_{n}({\bf r})$ is then
\[s_{n}({\bf r})=\left\{ \begin{array}{ll}
1 & \mbox{if ${\bf r}={\bf R}_{l}+{\bf u}_{l}^{n}$}\\
0 & \mbox{otherwise}\, . \end{array} \right. \]

The Fourier transform
\[{\bf u}({\bf q},k)=\sum_{n,l}{\bf u}_{l}^{n} \exp (i {\bf q} {\bf
R}_{l}+i k n d)\] identifies longitudinal $u^{l}({\bf q},k)={\bf
q} \cdot {\bf u}({\bf q},k)/q$ and transverse $u^{tr}({\bf
q},k)=[{\bf q} \times\hat{z}] \cdot {\bf u}({\bf q},k)/q$
components of ${\bf u}({\bf q},k)$
\[{\bf u}({\bf q},k)=u^{l}({\bf q},k){\bf q}/q+u^{tr}({\bf q},k)[{\bf
q} \times\hat{z}]/q  \, .\] where $\hat{z}$ is a unit vector in
the $z$ direction. The inverse transform is
\begin{equation}
 \uu_{l}^{n}=a^{2}d \int_{q,k} \uu(\qq,k) e^{-i\qq \RR_{l}-i k n d}
\end{equation}
where $\int_{q,k}=\int_{BZ}d^{2}q d k/(2\pi)^3$, $\int_{BZ}$ is
the ${\bf q}$ integration over the Brillouin zone (of area
$4\pi^{2}/a^{2}$) while $|k|<\pi/d$, and $a^2$ is the area of the
flux lattice's unit cell. Note that $\theta({\bf q},k)$ involves
much higher $q$ components, up to the cutoff $1/\xi$.

The next step is to consider small displacements in $F_v$, Eq.
(\ref{Fv}); this is equivalent to assuming that temperature is
well below the melting temperature $T_m$. This expansion
identifies the magnetic contribution to the elastic constants,
leading to a Hamiltonian of the form

\begin{eqnarray}
{\cal H}^{(2)} &&=\frac{1}{2}(d a^{2})^{2}\int_{q,k}
 \left  [q^{2} c_{66}^{0} + k_{z}^{2} c_{44}^{0}(k)\right]
   |u^{tr}({\bf q},k)|^{2} \nonumber\\
&&+{\cal H}_{el}\{u^l(\qq,k)\}+\frac{1}{2}\int ^{1/\xi}\frac{d^{2}
{\bf q} dk}{(2\pi)^{3}} G_{f}^{-1}(q,k)
|\theta({\bf q},k)|^{2}   \nonumber\\
&&-({\tilde J}/\xi^{2})\sum_{n}\int d^{2}\rr \cos\left (
\theta_{n}(\rr)+
\sum_{l}[\alpha(\rr-\RR_{l}-\uu_{l}^{n})-\alpha(\rr-\RR_{l}-\uu_{l}^{n+1})]
\right ) \label{H2}
\end{eqnarray}
where ${\cal H}_{el}\{u^l(\qq,k)\}$ involves elastic constants of
the longitudinal modes and
$k_{z}^{2}=\frac{4}{d^{2}}\sin^{2}\frac{kd}{2}$. The elastic
constants for the transverse modes, excluding Josephson coupling
terms, are given by \cite{glazman2,goldin}
\begin{eqnarray} c_{44}^{0}(k)&=&\frac{\tau}{8d a^2\lambda_{a
b}^2}\frac{1}{k_{z}^{2}}\ln\frac{1+k_{z}^{2}/Q_0^{2}}
{1+\xi^{2}k_{z}^{2}}\label{c44}\\
    c_{66}^{0}&=&\frac{\tau}{16da^2}\label{c66}
\end{eqnarray}
where  $Q_0^{2}=4\pi B/\Phi_{0}=4 \pi/a^{2}$ i.e. $\pi Q_0^2$ is
the area of a Brillouin zone, and
\begin{equation}\label{tau}
\tau=\frac{\Phi_{0}^{2}d}{4\pi^{2}\lambda_{a b}^{2}} \,.
\end{equation}
  If the Josephson term is expanded it would add
more terms to the elastic constants, however, near decoupling the
nonlinearity of the $\cos$ term is essential.

We can estimate the condition $T\ll T_m$ by evaluating the melting
temperature via the Lindeman criterion $\langle
(u_l^n)^2\rangle_0\approx c_L^2a^2$ where $c_L$ is a Lindeman
number of order $0.15$ and the average is with respect to the
elastic terms. The average is dominated by the softer transverse
modes yielding $T_m \ln(2\pi ^2 \lambda_{ab}^2/a^2)\approx c_L^2
\tau$, i.e $\tau$ is the temperature scale for melting. Melting in
the absence of Josephson coupling was in fact studied
\cite{dodgson2}, showing that $T_m$ is between $\tau/8$ and the
two-dimensional melting temperature $\approx 0.004\tau$,
approaching the latter at high fields $a\ll \lambda_{ab}$. The
condition $T_d\ll T_m$ is in fact satisfied only at $a\ll
\lambda_{ab}$ [Eqs. (\ref{Td00},\ref{tds},\ref{tdb}) below], hence we limit
our solutions to $\lesssim 0.01\tau$.

The next simplification, in accord with the condition of small
 displacements $|u_l^n|\ll a$ is an expansion of the relative
 phase in the Josephson term.
\begin{equation}
\sum_{l}[\alpha(\rr-\RR_{l}-\uu_{l}^{n})-\alpha(\rr-\RR_{l}-\uu_{l}^{n+1})]
\simeq \sum_{l}(\uu_{l}^{n+1}-\uu_{l}^{n})\cdot {\mbox{\boldmath
$\nabla$}} \alpha({\bf r}-{\bf R}_{l}^{n}) \equiv {\tilde
b}_{n}(\rr)\label{tbn}
\end{equation}

Since ${\tilde b}_{n}(\rr)$ contains all $\qq$ in its Fourier
transform it is useful to separate from it the components within
the 1st Brilluin zone,
 \begin{eqnarray}\label{bn}
{\tilde b}_{n}(\rr)&=& b_{n}(\rr) + B_{n}(\rr)\nonumber\\
b_{n}(\rr)&=&2\pi i d \int_{BZ}\frac{d^{2}\qq d
k}{(2\pi)^{3}}e^{-i
\qq \cdot \rr-i k n d} (e^{i k d}-1)\frac{u^{tr}(\qq,k)}{\qq} \nonumber\\
 B_{n}(\rr)&=& 2\pi i d \sum_{\QQ\neq 0}
\int_{BZ}\frac{d^{2}\qq d k}{(2\pi)^{3}}e^{-i (\qq+\QQ)\cdot \rr-i
k n d} (e^{i k d}-1)\frac{\uu(\qq,k)\cdot [\hat{z}\times
(\qq+\QQ)]}{(\qq+\QQ)^{2}}
\end{eqnarray}
The last term displays the Fourier transform of ${\mbox{\boldmath
$\nabla$}}\alpha(\rr)$, which for $Q=0$ projects the transverse
displacement.

The lowest order effect of thermal fluctuations is obtained as an
average with respect to the elastic terms, i.e.
\begin{equation}\label{av}
\langle \cos[\theta_{n}(\rr)+{\tilde
b}_{n}(\rr)]\rangle_0=\exp[-\frac{1}{2}
\langle\theta_{n}^2(\rr)\rangle_0-\frac{1}{2} \langle {\tilde b}
_{n}^2(\rr)\rangle_0]
\end{equation}
The only singular term in this average is due to $b_n(\rr)$ from
Eq. (\ref{bn}) which in fact drives the decoupling transition,
i.e. even if the fluctuations in $u_l^n$ are small, their effect
on the Josephson phase can be divergent. The average $\langle
B_n^2(\rr)\rangle_0$ is readily shown to yield $\approx T/\tau$,
up to logarithmic terms, and since $T\ll \tau$ below melting its
effect in Eq.(\ref{av}) is negligible.

We wish to integrate out also the high $\qq$ modes of
  $\theta(\qq,k)$, i.e. momenta in the range $Q_0<q<1/\xi$. The
  effect from Eq. (\ref{av}) is a factor
\[ D=e^{-\frac{1}{2} \int_{1/a}^{1/\xi} d^{2}  q d k
G_{f}(q,k)/(2\pi)^{3}} =e^{-\frac{T}{\tau} \ln \frac{a}{\xi} } \]
which can also be neglected.

The effective Hamiltonian is now defined with a momentum cutoff
$q<Q_0$ and the corresponding smallest length scale is $a$. The
Josephson term involves $J\sum_r \rightarrow (J/a^2)\int d^2r$ so
that $J$ is now the Josephson coupling energy per area $a^2$, i.e.
$J={\tilde J}(a/\xi)^2$. Since $b_n(\rr)$ depends only on the
transverse modes all longitudinal terms are decoupled and we can
consider an effective Hamiltonian for the transverse modes
\begin{eqnarray}
{\cal H}^{(3)} &=&\frac{1}{2}\sum_{\qq,k} \left(G_{f}^{-1}(q,k)
|\theta({\bf q},k)|^{2} +G_{b}^{-1}(q,k)|b(\qq,k)|^{2} \right)
\nonumber\\
&- &(J/a^{2})\sum_{n}\int d^{2}\rr \cos (
\theta_{n}(\rr)+b_{n}(\rr))\label{H3}
\end{eqnarray}
where
\begin{equation}
G_{b}^{-1}(q,k)=q^{2}\frac{ a^{4}}{(2\pi d)^{2}}\left[
c_{44}^{0}+q^{2}c_{66}^{0}/k_{z}^{2}\right] \,.
\end{equation}

Finally we shift
\[\phi_{n}(\rr)=b_{n}(\rr)+\theta_{n}(\rr)\]
so that $\theta(\qq,k)$ can be integrated out leading to a
Gaussian term $\frac{1}{2}\sum_{\qq,k}[G_{b}(q,k)
+G_{f}(q,k)]^{-1} |\phi(\qq,k)|^{2}$. It is readily shown that
$G_{f}(q,k)/G_{b}(q,k)\ll 1$ for all $\qq$, $k$, except when both
 $k<1/\lambda_{ab}$ and  $q\gg k\lambda_{ab}/a$ which  has negligible
effect since large $k$ dominates the following integrals.
Therefore we neglect $G_{f}(q,k)$ and our final Hamiltonian is
\begin{equation}\label{H}
{\cal H} =\frac{1}{2}\sum_{\qq,k}G_{b}^{-1}(q,k) |\phi(\qq,k)|^{2}
-(J/a^{2})\sum_{n}\int d^{2}\rr \cos \phi_{n}(\rr)
\end{equation}
Our task is then to evaluate the partition sum
\begin{equation}\label{Z2}
{\cal Z}=\int {\cal D} \phi e^{-{\cal H}\{\phi\}/T}\,.
\end{equation}

\section{General properties}

In this section we identify the order parameter of the decoupling transition and related correlation functions. Also a scheme for evaluating the frequency of the Josephson plasma resonance is given.

In the RG procedure, as detailed in section IV, a phase transition is found at a temperature $T_d$ such that at $T>T_d$ $J$ is irrelevant (scales to zero under RG) while at $T<T_d$ it is relevant (scales to strong coupling). To identify the order parameter of this transition we consider first the Hamiltonian Eq. (\ref{H}) in presence of an external vector potential in the $z$ direction $A_n(\rr)$ which is $z$ independent between layers. The relative superconducting phase on neighboring layers $\phi_n(\rr)$ is then shifted by $(2\pi d/\phi_0)A_n(\rr)$ and the Hamiltonian becomes
\begin{equation}\label{Hgauge}
{\cal H_A} =\frac{1}{2}\sum_{\qq,k}G_{b}^{-1}(q,k) |\phi(\qq,k)|^{2}
-(J/a^{2})\sum_{n}\int d^{2}\rr \cos [\phi_{n}(\rr)-\frac{2\pi d}{\phi_0}A_n(\rr)]\,.
\end{equation}
The Josephson current is a derivative of the free energy ${\cal F}=-T\ln {\cal Z}$
\begin{equation}\label{jz}
j_z(\rr,n)=-\frac{c}{d}\frac{\partial {\cal F}}{\partial A_n(\rr)}=\frac{2\pi c}{\phi_0}\frac{J}{a^2} \langle \sin [\phi_{n}(\rr)-\frac{2\pi d}{\phi_0}A_n(\rr)]\rangle _A\,.
\end{equation}
The linear response to $A_n(\rr)$ has a $\sim \cos \phi_{n}(\rr)$ term as well as a nonlocal term from the expansion of ${\cal H_A}$ in $\exp (-{\cal H_A}/T)$,
\begin{equation}
j_z(\rr,n)=-\left(\frac{2\pi}{\phi_0}\right)^2\frac{Jdc}{a^2}[\langle \cos \phi_{n}(\rr) \rangle A_n(\rr)-\frac{J}{Ta^2}\sum_{n'}\int d^2r' \langle \sin \phi_{n}(\rr) \sin \phi_{n'}(\rr')\rangle A_{n'}(\rr')]
\end{equation}
where averages $\langle ...\rangle$ are in the $A=0$ system.
The superconducting response is identified by a uniform $A_n(\rr)=A$ so that the superconducting phase $\varphi_n(\rr)$ acquires a uniform twist $\varphi_n(\rr)-(2\pi /\phi_0)An$. Note that the partition sum excludes $\varphi_n(\rr)$ with a global twist, hence the $A$ term cannot be transformed away. The superconducting response $Q$ is defined by $j_n(\rr)=QA$, i.e.
\begin{equation}\label{Q}
Q=\left(\frac{2\pi}{\phi_0}\right)^2\frac{Jdc}{a^2}[\langle \cos \phi \rangle-\frac{J}{Ta^2}\sum_{n}\int d^2r \langle \sin \phi_{0}(0) \sin \phi_{n}(\rr)\rangle ]
\end{equation}
where $\langle \cos \phi_{n}(\rr) \rangle$ is $n$ and $\rr$ independent is written for short as
$\langle \cos \phi  \rangle$.
This order parameter was identified by Li and Teitel \cite{teitel} as a helicity modulus and by Koshelev \cite{koshelev} as the superconducting response (i.e. the zero frequency limit of his Eq. 3). This order parameter signifies breaking of gauge symmetry: at $T>T_d$ the irrelevancy of $J$ implies diverging fluctuations in $\phi_n(\rr)$, hence $A_n(\rr)$ has no effect on $Z$ and $Q=0$. At $T<T_d$ $J$ flows to strong coupling so that $\phi_n(\rr)$ has finite fluctuations near the energy minimum where $\phi_n(\rr)=0$; this implies that a gauge transformation of $A_n(\rr)$ must be combined with a change in $\phi_n(\rr)$. The manifestation of this broken gauge symmetry is
 $Q\neq 0$ or a finite Josephson current.

The Josephson plasma resonance frequency $\omega_{pl}$ is a significant probe \cite{shibauchi,matsuda} of a possible decoupling transition.
As shown by Koshelev \cite{koshelev} the superconducting response at $\omega_{pl}$ is dominated by the 1st term of Eq. (\ref{Q}), so that
$\omega_{pl}\sim \langle \cos \phi \rangle$. We reconsider this relation in section V while here we present a scheme for evaluating $\langle \cos \phi \rangle$. We note first that a derivative of the free energy yields
\begin{equation}
\langle \cos \phi \rangle= \frac{a^2}{NL^2}\frac{\partial}{\partial J}{\cal F}(J)
\end{equation}
where $L^2$ is a layer area and $N$ is the number of layers. While the details of the RG procedure are not needed in this section, some general properties can be derived from the asymptotic RG transformation. A general RG changes the original lattice unit $a$ to a renormalized one $a^R$ while $J(a)\rightarrow J(a^R)=J^R$; additional parameters in ${\cal F}(J)$ (not displayed here) may also be renormalized while ${\cal F}$ itself changes by
\begin{equation}\label{dF}
d{\cal F}=-NL^2f(a)da
\end{equation}
In 1st order RG $f(a)=0$ (section IVA) but becomes finite in 2nd order (section IVB) so that $f(a)\sim J^2(a)$. Hence integrating Eq. (\ref{dF}) yields a J dependent term for either a relevant or irrelevant $y$, hence $\langle \cos \phi \rangle$ is finite in either the coupled or the decoupled phases. We consider in particular the decoupled phase $T>T_d$ and a system size $L\gg a$. At the final stage of RG when $a^R\rightarrow L$ the renormalized $J^R$ becomes extremely small and one can safely use 1st order RG which has the form (section IVA)
\begin{equation}\label{RG0}
dJ/J=2(1-Z)da/a
\end{equation}
where $Z=Z(J,T)$ contains renormalizations due to higher order RG terms at shorter scales; note that $Z=1$ at $T=T_d$ since for $Z>1$ Eq. (\ref{RG0}) shows that $J^R\sim L^{2(1-Z)}\rightarrow 0$ at $L\rightarrow \infty$. E.g. if 1st order RG is used all the way from the initial values then $Z=T/T_d^0$ where $T_d^0$ is Eq. (\ref{Td0}) below. Integrating Eq. (\ref{dF}) yields
\begin{equation}
F(J^R)-F(J)=-NL^2\int_a^{a^R} f(a')da'  \,.
\end{equation}
At the scale $a^R=L$ the system has just one degree of freedom so that the term $\sim q^2\phi(\qq,k)$ in Eq. (\ref{H}) is absent, hence
\begin{equation}
F(J^R)=-T\ln [\int_0^{2\pi}d\phi e^{J^R\cos \phi/T}]^N=-\frac{1}{4}TN[(\frac{J^R}{T})^2+O(\frac{J^R}{T})^4] \,.
\end{equation}
Thus $F(J^R)\sim L^{4(1-Z)}$ at $T>T_d$ so that at $L\rightarrow \infty$
\begin{equation}\label{cosRG}
\langle \cos \phi \rangle=a^2\frac{\partial}{\partial J}\int_a^{\infty}f(a')da' +O(L^{2-4Z}) \,.
\end{equation}
 RG provides then an efficient method for evaluating $\omega_{pl}$ via (\ref{cosRG}).

We proceed now to study correlations. We note first the correction in Eq. (\ref{cosRG})
$\langle \cos \phi \rangle_L-\langle \cos \phi \rangle\sim L^{2-4Z}$ (the upper limit in $\int^Lf(a')da'$ gives the same exponent as can be seen from section IV). This correction determines the decay rate of the correlation function, by considering the 2nd derivative $\partial^2{\cal F}/\partial J^2\sim (\partial /\partial J)\int {\cal D}\phi \cos \phi_0(0) e^{-H}/{\cal Z}$. This involves a $\cos \phi_n(\rr)$ correlation as well as $\langle \cos \phi \rangle^2$ from $\partial {\cal Z}/\partial J$, hence
\begin{equation}\label{ddJ}
\frac{\partial \langle \cos \phi \rangle}{\partial J}=\frac{1}{Ta^2}\sum_n\int d^2r \langle \cos \phi_0(0) \cos \phi_n(\rr)\rangle_c
\end{equation}
where $\langle \cos \phi_0(0) \cos \phi_n(\rr)\rangle_c=\langle \cos \phi_0(0) \cos \phi_n(\rr)\rangle -\langle \cos \phi \rangle^2$ vanishes at large $\rr$.
To reproduce the finite size correction $L^{2-4Z}$ at $T>T_d$ the correlation must decay as $\langle \cos \phi_0(0) \cos \phi_0(\rr)\rangle_c \sim 1/r^{4Z}$.

In the coupled phase $a$ approaches the correlation length $\xi_d$ at which $J/T\approx 1$ becomes a strong coupling. The Hamiltonian can then be expanded as ${\cal H}\sim \sum_q(q^2+\xi_d^{-2})|\phi(\qq,k)|^2$, keeping just the $q\rightarrow 0$ form. Hence for $r>\xi_d$ the system becomes Gaussian and the correlations decay exponentially. We conclude that
\begin{eqnarray}\label{Jcor2}
\langle \cos \phi_0(0) \cos \phi_0(\rr)\rangle_c&\sim& 1/r^{4Z} \qquad T>T_d \nonumber\\
    &\sim& e^{-r/\xi_d} \qquad T<T_d \,.
\end{eqnarray}
This distinctive behavior of the correlations serves as an additional identification of the phase transition.

We finally examine the validity of the high temperature expansion which provides an easy estimate of $\langle \cos \phi\rangle$. We define $\langle ...\rangle_0$ as an average with respect to the $J=0$ system, so that
\begin{equation}\label{bb}
\langle [\phi_n(0)-\phi_n(\rr)]^2\rangle_0 =T\frac{4\pi^2d^2}{a^4}\int
\frac{dk}{2\pi c_{44}^0(k)}\int \frac{d^2q(1-e^{i\qq\cdot\rr})}{4\pi^2
q^2(1+q^2/q_u^2)}
\end{equation}
where the $c_{66}$ term is absorbed into $q_u^2=4\ln
(a/d\sqrt{\pi})/\lambda_{ab}^2$ (at the dominant $k=\pi/d$ where $c_{44}^0$ is significantly softer). The k integration yields the 1st order $T_d^0$ (Eq. \ref{Td0} below) so that
\begin{eqnarray}\label{bcor}
\langle [\phi_n(0)-\phi_n(\rr)]^2\rangle_0=4t\int \frac{d^2q}{\pi q^2(1+q^2/q_u^2)}(1-e^{-i\qq\cdot\rr})&=
& 8t\ln (rq_u) \qquad r>1/q_u  \nonumber\\
&=& -2t\,r^2q_u^2\ln (rq_u) \qquad \qquad    r<1/q_u
\end{eqnarray}
where $t=T/T_d^0$. Using the Gaussian average
\begin{equation}\label{Jcor3}
\langle \cos \phi_n(0)\cos \phi_n(\rr ) \rangle_0 =\exp \{-\langle [\phi_n(0)-\phi_n(\rr)]^2\rangle_0/2 \}
\end{equation}
and expansion of ${\cal Z}$ to 1st order in $J/T$ we obtain
\begin{equation}\label{av1}
\langle \cos \phi\rangle= \frac{J}{Ta^2}\int d^2r'\langle \cos
\phi_n(0)\cos \phi_n(\rr ) \rangle_0 \approx \frac{\pi J}{Ta^2q_u^2(2t-1)}  \,;
\end{equation}
the contribution of $r<1/q_u$ which becomes comparable at large $t$ is omitted. To check the validity of (\ref{av1}) we note that it can also be obtained from the exact relation (\ref{ddJ}) taken as a perturbation in $J$,  where $\langle \cos \phi \rangle \sim J^2$ is formally of higher order. However, this term is actually $\sim J^2L^2N$, hence the perturbation expansion formally fails at $L, N \rightarrow \infty$.

We note that within this naive perturbation expansion the correlation decays (incorrectly) to zero,
\begin{equation}\label{Jcor1}
<\cos\phi^{n}(\rr)\cos\phi^{n}(0)>\sim  (rq_u)^{-4t}
+O[(J/T)^2(rq_u)^{2-4t}]
\end{equation}
with a finite contribution to $\langle \cos \phi\rangle$ in Eq. (\ref{av1}). We note also that the next order terms in $J$, while decaying more slowly, are still convergent in Eq. (\ref{av1}) for $t>1$. In fact the result (\ref{av1}), quiet remarkably, reproduces the 2nd order RG result (up to a numerical prefactor) as found below. The reason is that the decay of (\ref{Jcor1}) is actually correct for $\langle \cos \phi_0(0) \cos \phi_n(\rr)\rangle_c$ as in Eq. (\ref{Jcor2}) (with $Z=t$ in weak coupling), which when substituted in (\ref{ddJ}) reproduces the form (\ref{av1}).

We summarize the salient features of the decoupling transition: (i) Relevancy of $J$ corresponds to a broken gauge symmetry in the coupled phase, (ii) an order parameter that corresponds to (i) is $Q$ of Eq. (\ref{Q}) or the Josephson current, and (iii) decay of correlations, such as $\langle \cos \phi_0(0) \cos \phi_n(\rr)\rangle_c$, is either exponential in the coupled phase or a power law in the decoupled phase.

\section{RG solution}

The decoupling transition is driven by the singular response of
the Josephson phase $b_n(\rr)$ in Eq. (\ref{bn}). The singularity
is due to the long range effect of an individual shift $u_l^n$
which decays as ${\bf \nabla} \alpha({\bf r}-{\bf R}_{l}^{n}) \sim
[{\bf r}-{\bf R}_{l}^{n}]^{-1}$ (see Eq. \ref{tbn}). The
contributions of many such small displacements at a given location
$\rr$ result in a divergent response at the decoupling transition. The RG method is designed to handle such divergences and avoid the pitfalls of naive perturbation expansions.

\subsection{1st order RG}

 The RG method proceeds by integrating out slices of
high momentum shells $\Lambda-d\Lambda<q<\Lambda$, and the
momentum cutoff $\Lambda$ is successively reduced; initially
$\Lambda=Q_0$. The field $\phi(\qq,k)$ is then decomposed as
$\phi(\qq,k)=\zeta(\qq,k)+\chi(\qq,k)$ where $\zeta(\qq,k)$
carries momenta in the range $\Lambda-d\Lambda<q<\Lambda$ while
$\chi(\qq,k)$ has momenta $q<\Lambda-d\Lambda$. In this subsection
we analyze the phase transition by using the 1st order equation in
J. Expansion of Eq. (\ref{Z2}) to 1st order in $J$ and averaging
on the high momentum shell yields
\begin{equation}
<\cos(\zeta_{n}(\rr)+\chi_{n}(\rr))>_{\zeta}=\cos\chi_{n}(\rr)
(1-\frac{1}{2}<(\zeta_{n}(\rr))^{2}>)\nonumber\,.
\end{equation}
Defining \[G_b^{-1}(q,k)=\frac{q^{2}}{8\pi d}\frac{T}{g(q,k)} \]
we obtain $<(\zeta_{n }(\rr))^{2}>=d\int
\frac{dk}{2\pi}g(\Lambda,k)d\Lambda/\Lambda$. Rescaling
$a\rightarrow a+da$ ($da/a=d\Lambda/\Lambda$) leads to 1st order
RG, i.e. it identifies the change in the coefficient $y$ of the $\cos$
term, with the initial value $y_0=J/T$, as
\begin{equation}\label{yeq}
\frac{dy}{y}=[1-d\int \frac{dk}{2\pi}g(\Lambda(x),k)]\frac{dx}{x}
\end{equation}
where $x=Q_0^2/\Lambda^2$ is in the range $1<x<\infty$. For a
continuous transition the limiting
 $g(\Lambda\rightarrow 0,k)$ can be taken and the vanishing of the
 right hand side in Eq. (\ref{yeq}) identifies the decoupling
 transition temperature (which is independent of $y_0$)
\begin{equation}\label{Td0}
T_d^0=\frac{4a^4}{d^2}\left[ \int
\frac{dk}{c_{44}^0(k)}\right]^{-1}\,.
\end{equation}
This defines the units of our temperature variable $t$,
\begin{equation}
t=T/T_d^0
\end{equation}
and Eq. (\ref{yeq}) has then the form (\ref{RG0}) with $Z=t$.

The asymptotic solution of Eq. (\ref{yeq}) is $y(x)=y_0x^{1-t}$.
For $t>1$ the asymptotic $y(x)$ vanishes on
long scales $x\rightarrow \infty$, hence the meaning of decoupling
is that the Josephson coupling vanishes on long scales. For $t<1$
the coupling $y(x)$ increases, RG stops then when $x$ reaches
$x_d=(\xi_d/a)^2$ where strong coupling $y(x)\approx 1$ is
achieved; this identifies a correlation length
\begin{equation}\label{xid}
\xi_d\approx a (y_0)^{1/[2(t-1)]}\,.
\end{equation}

An explicit form for $T_d^0$ can
 be derived by noting the significant softening of
 $c^{0}_{44}(k)$ at $k>1/a$
which implies that the $k$ integration in $g(\Lambda,k)$ is
dominated by $k\approx \pi/d$. Hence we replace in Eq. (\ref{c44})
$\ln [(1+k_{z}^{2}/Q_{0}^{2})/ (1+\xi^{2}k_{z}^{2})]\rightarrow
2\ln (a/d\sqrt{\pi})$, resulting in
\begin{eqnarray}
G_b(q(x),k)&=&da^2\frac{4xt\sin^2(kd/2)}{T(1+\frac{1}{4gx})}
\label{Gb1}\\
g(q(x),k)&=& \frac{2t\sin^2
(kd/2)}{1+\frac{1}{4gx}}\label{g1}
\end{eqnarray}
so that
\begin{equation}\label{intg}
d\int \frac{dk}{2\pi}g(\Lambda(x),k)=\frac{t}{1+1/(4gx)}\,.
\end{equation}
$T_d^0$ and the parameter \cite{dodgson} $g$ are
\begin{eqnarray}\label{Td00}
g&=&\frac{a^{2}}{4\pi \lambda_{a
b}^{2}}\ln\frac{a}{d\sqrt{\pi}}\nonumber\\
T_d^0&=&g\tau=\frac{\tau a^{2}\ln(a/d\sqrt{\pi})}{4\pi
\lambda_{a b}^{2}}\,.
\end{eqnarray}
Note that $1/4gx=q^2/q_u^2$ where $q_u$ is defined below
Eq. (\ref{bb}).
 A continuous transition is determined by the
$x\rightarrow \infty$ behavior, hence Eq. (\ref{intg})$\rightarrow
t$ and the critical point is at $t=1$. We will examine below the
possibility of a 1st order transition, hence in general we keep
$\sim 1/x$ terms. The solution for Eq. (\ref{yeq}) is then
\begin{equation}\label{yx1}
y(x)= y_{0} x\left(\frac{1+4g}{1+4gx}\right)^t
\end{equation}
and the phase transition is indeed at $t=1$.

\subsection{2nd order RG}

We proceed to study 2nd order RG. Our main objective in this
subsection is to see if RG can reproduce a 1st order transition as
proposed within the SCHA method \cite{daemen}.

The RG procedure for a Hamiltonian of the type (\ref{H}) has
been derived in appendix A of Ref \cite{h3} up to 2nd order in
$J$. In this process new terms in the effective Hamiltonian are
generated so that the partition sum has the form
\begin{equation}\label{Z1}
Z=\int {\cal D} \phi \exp \left[ -\frac{1}{2} \int \frac{d^{2}\qq
d k}{(2\pi)^{3}}G_R^{- 1}(\qq,k)|\phi(\qq,k)|^{2}+\int \frac
{d^{2}\rr}{a^{2}} \{y \cos[\phi_{n}(\rr)]+v
\cos[\phi_{n}(\rr)+\phi_{n+1}(\rr)] \}\right]
\end{equation}
where we define
\begin{equation}\label{GR}
G^{-1}_{R}(q,k)=\frac{q^{2}}{8\pi d}\left(
\frac{1}{g(q,k)}+h_{0}+h_{1}\cos kd
      \right)\,.
\end{equation}

The new variables generated by RG, $v(x)$, $h_{0}(x)$, $h_{1}(x)$,
are cutoff dependent with initial values of
$v_0(1)=h_0(1)=h_1(1)=0$. Note that the renormalization of $h^{0}$
and $h^{1}$ is equivalent to renormalization of $c^{0}_{44}$ . The
recursion relations to 2nd order in $y$ are \cite{h3}
\begin{eqnarray}
        d y&=&[ y (1-X_{0})+ y v(X_{0}+X_{1})]d x/x \nonumber\\
d h_{0}&=&[ y^{2}X_{0}+4 v^{2}(X_{0}+X_{1})]d x/x \nonumber\\
         dv&=&[ v (1-2 X_{0}-2 X_{1})-\frac{1}{4} y^{2} X_{1}]d x/ x
\nonumber\\
 dh_{1}&=&[4 v^{2}(X_{0}+X_{1})]d x/x \label{RG1}
\end{eqnarray}
where $X_{n}$, $n=0,1$, are $h_{0}$, $h_{1}$ and $x$ dependent,
\begin{equation}
X_{n}=d\int \frac{d k}{2\pi}\frac{\cos(nkd)}{1/g(x,k)
+h_{0}+h_{1}cos k d} \label{X01} \,.
\end{equation}
We have absorbed factors $\gamma, \gamma'$ of order $1$ in the definitions of $y,v$ which depend on the cutoff smoothing procedure \cite{h3}; the initial value of $y$ is then $y(1)=\gamma' y$.
 The equations (\ref{RG1}) are to be integrated from their
initial values.

 To first order in $y$ we rederive Eq. (\ref{yeq}) above,
\begin{equation}
d \ln y=[  1-X_{0}(x, h_0=h_1=0)]d x/x\nonumber
\end{equation}

Before presenting numerical solutions for Eqs. (\ref{RG1}), it is
instructive to consider a simple situation where $v$ and $h_{1}$
are neglected. This is a reasonable approximation since we
eventually find that $v<y$ and $h_{1}<h_0$. The RG equations are
then
\begin{eqnarray}
d y&=&y (1-X_{0}(x))d x/x \nonumber\\
d h_{0}&=&y^{2}X_{0}(x)d x/x \label{RG2}
\end{eqnarray}
where [using the approximate form for the $\ln$, as above Eq.
\ref{Gb1}]
\begin{equation}\label{X0}
X_0(x)=d\int
\frac{dk}{2\pi}\frac{1}{\frac{1+1/(4gx)}{2t\sin^2(kd/2)}+h_0(x)}\,.
\end{equation}

 We consider first the asymptotic solution at $x\rightarrow
\infty$ in the regime where $y$ is relevant, i.e. the coupled
phase. Integrating the 1st equation of (\ref{RG2}) we obtain
\begin{equation}
\ln \frac{y}{y_0}=\ln x - \int_1^x X_0(x')\frac
{dx'}{x'}\,.\nonumber
\end{equation}
We claim that the 2nd term converges, hence the parameter $s$
\begin{equation}\label{s}
s=\int_1^{\infty}dx \frac{X_0(x)}{x}=d\int
\frac{dk}{2\pi}\int_1^{\infty}dx\frac{1}{\frac{x+1/4g}{2t\sin^2(kd/2)}+xh_0(x)}
\end{equation}
defines the asymptotic form of $y$,
\begin{equation}\label{yasym}
y=y_0xe^{-s}\,.
\end{equation}
The RG equations are valid only up to $y\approx 1$, i.e. for a
small $y_0$ up to a large $x_d=(\xi_d/a)^2$, but not infinite. Eq.
(\ref{s}) therefore assumes that by formally extending Eq.
(\ref{RG2}) to $x\rightarrow \infty$ the integration range between
$\xi_d$ and $\infty$ is negligible.

We note that the $h_0(x)$ term represents an additional mass term
in the propagator $G_R(q,k)$, Eq. (\ref{GR}) if $h_0(x)\sim x$.
The $xh_0(x)$ term in Eq. (\ref{s}) is then $\sim x^2$ and the
integral is convergent. In the SCHA method such a mass term serves
as a variational parameter and an analogous equation to (\ref{s})
is derived in the next section. The essential point here is that
$xh_0(x)$ is $\sim x^2$ only asymptotically while its scale
dependence at finite $x$ can be different, resulting in a
different critical behavior. The scale dependence of the "mass"
$h_0(x)/x$ is a feature which is beyond either 1st order RG or SCHA.

To complete the argument we need to show that $h_0(x)$ increases
with $x$ justifying the convergence of (\ref{s}). Assuming that
$h_0(x)$ increases with $x$, we can use $X_0\sim 1/h_0$ to yield
from Eqs. (\ref{RG2},\ref{yasym}) that indeed $h_0\approx y \sim
x$. [For $x<x_d$ more terms in the the denominator of Eq.
(\ref{X0}) need to be kept, modifying the way $h_0$ increases with
$x$.]

We proceed now to solve the RG equations (\ref {RG2}) in general form.
Integrating Eq. (\ref{X0}) we obtain an explicit form
\begin{equation}\label{X02}
 X_{0}=\frac{t}{
1+ t h_{0}+\frac{1}{4g x} }\cdot f(h_{0},x,t)
\end{equation}
where
\[f(h_{0},x,t)=\frac{1}{b}\left(1-\sqrt{\frac{1-b}{1+b}}\right)\]
and \(b=t h_{0}/(1+1/(4gx)+h_{0}t)\). The function
$f(h_{0},x,t)$ varies slowly between \mbox {$0.84 <f(h_{0},x,t)<1$}.

The RG Eqs. (\ref{RG2}) for $y$ and $h_{0}$ become
\begin{eqnarray}
        dy&=&y(1-t \frac{f(h_{0},x,t)}{(1+t h_{0}+\frac{1}{4g
x})}\,\frac{dx}{x}\nonumber\\
dh_{0}&=&t\frac{y^{2}f(h_{0},x,t)}{1+ t
h_{0}+\frac{1}{4g x}}\,\frac{dx}{x} \label{yh0}
\end{eqnarray}

\begin{figure}
\begin{center}
\includegraphics[scale=0.7]{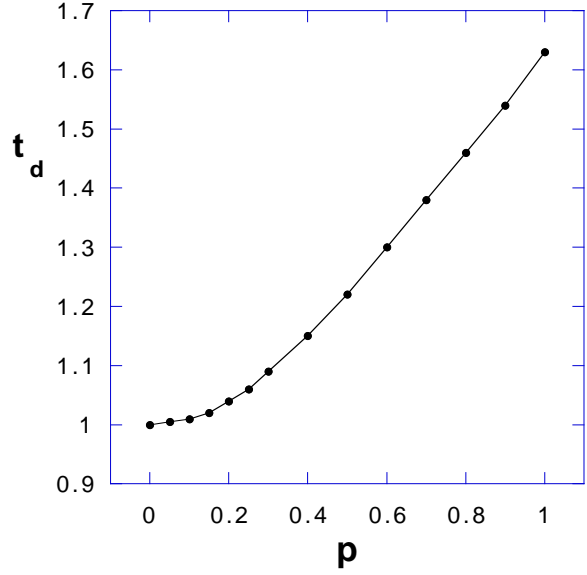}
\end{center}
\caption{The decoupling temperature $t_{d}(p)=T_{d}(p)/T_d^0$
derived from the numerical solution of Eq. (\ref{yh0}) with
$4g=0.0001$.}
\end{figure}

If the $1/(4g x)$ term in Eq. (\ref{yh0}) is neglected and
$f(h_0,x,t)=1$ is taken, Eq. (\ref{yh0}) is equivalent to the
Kosterlitz-Thouless equations \cite{h3} which show that the
critical temperature is enhanced by a factor $1+y_0$ and that
$y\sim x$ when $y$ is relevant. The presence of the
$1/4gx$ term leads, however, to a more significant
enhancement which is measured by a parameter $p$,
\begin{equation}\label{p}
p=\sqrt{\frac{Ty_0}{2\tau
g^2}}=\sqrt{\frac{ty_0}{2g}}
\end{equation}
measuring the ratio of two small parameters $y_0$ and
$g$. It is also useful to express $p$ in terms of the
Josephson length
$\lambda_J=\Phi_0a\sqrt{d}/(4\lambda_{ab}\sqrt{J\pi^3})$, i.e.
$p=(8\pi)^{-1/2}a/(g\lambda_J)$.

In appendix A we derive a form for the critical temperature which
yields a proper $p$ dependence for both small and large $p$,
\begin{eqnarray}
T_{d}&=&(1+2 p^{2})g\tau \qquad p\ll 1
\label{tds}\\
 T_{d}&=& \frac{3}{2} g\tau p
\qquad \qquad  p \gg 1       \label{tdb}
\end{eqnarray}

 We have solved Eq. (\ref{yh0}) numerically and the results for $T_d$
 are shown in Fig. 1. The $p$-dependence is indeed significant,
 becoming linear at high $p$; in the range $1\lesssim p <1000$ we
 find $t_d=0.95p$ . Thus $T_d\sim 1/B$ (Eq. \ref{Td00})
 at $p\ll 1$ while $T_d\approx (a/\lambda_J)\tau\sim 1/\sqrt{B}$
 at $p\gg 1$. We note also the significant enhancement in the
 value of $T_d$ in the latter case, i.e. $T_d\gg T_d^0$. We note that the RG expansion is valid for $y_0\ll 1$, which limits Eq. (\ref{tdb}) (by inserting it in Eq. (\ref{p})) to $1\ll p\ll 1/g$.

\begin{figure}
\begin{center}
\includegraphics[scale=0.7]{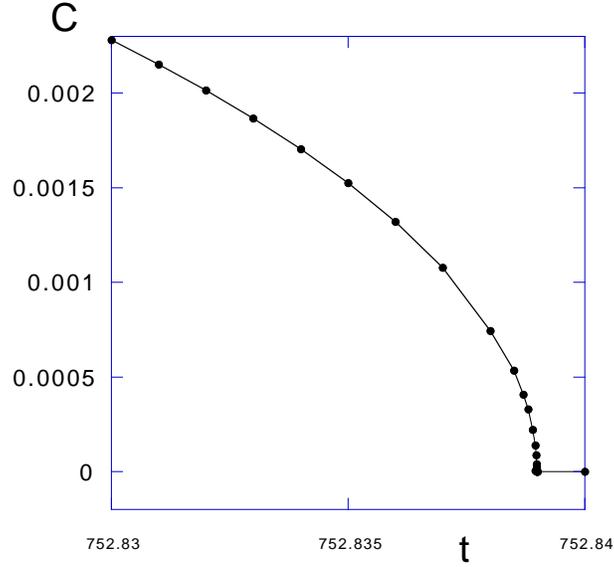}
\end{center}
\caption{The temperature dependence of the
 correlation length $\xi_d$ in terms of $C=[1+2\ln (\xi_d/a)]^{-1}$
 near the decoupling transition temperature
$t_{d}=T_{d}/T_d^0=752.839$. The numerical solution of Eq.
(\ref{yh0}) uses $4g=0.0001$ and  initial conditions $y_{0}=0.01$,
$h_{0}=0$, .}
\end{figure}

At $T<T_{d}$ we find that the variable $y$ first increases, then
decreases, corresponding to weakening of the $1/4gx$ term,
and finally increases up to $y\approx 1$ at the scale
$x=x_d=(\xi_d/a)^2$ where we should stop the renormalization
process. The temperature dependence of $x_d$ is shown on Fig 2. We
see that the correlation length $\xi_d \rightarrow \infty$ as
$T\rightarrow T_{d}$ which shows that the phase transition is a
continuous one. We note that the early estimate of Glazman and Koshelev \cite{glazman2} of the decoupling temperature gives a result similar to Eq. (\ref{tdb}). They derive a condition for large fluctuations $\langle\phi^2\rangle$ which by itself does not prove a phase transition; furthermore the estimated critical temperature vanishes with $p$, i.e. it is incorrect at $p\ll 1$. The large fluctuation condition is close in spirit to the SCHA method and is further discussed in section V.

We present now numerical solutions of the full RG equations
(\ref{RG1}) with $X_{0}$ and $X_{1}$ given by Eq.(\ref{X01}) in
Figs. 3, 4, for the same initial conditions as in Fig. 2
[$y_0=0.01$, $h_0(x=1)=0$] and
$v_{0}=0$, $h_{1}(x=1)=0$. We choose $4g=0.0001$ and these initial conditions
since in this case the SCHA method (see section IV) yields a 1st
order transition.

\begin{figure}
\begin{center}
\includegraphics[scale=0.7]{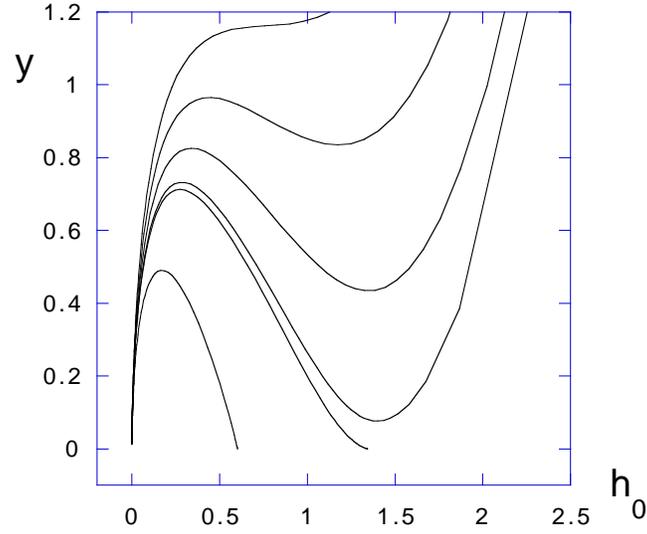}
\end{center}
\caption{The numerical solution of the RG Eqs.
(\ref{RG1}) with $4g=0.0001$ and initial conditions
$y_{0}=0.01$ and $h_0=h_1=0$,
 projected on the $y-h_{0}$ plane
 for different temperatures near the temperature of the decoupling
transition $t_{d}=T_{d}/T_d^0=825.7$. A higher curve corresponds
to a higher $t$, i.e. the two lower curves have $t<t_d$ while the
others have $t>t_d$. \vspace{0.5cm}}
\end{figure}

\begin{figure}
\begin{center}
\includegraphics[scale=0.7]{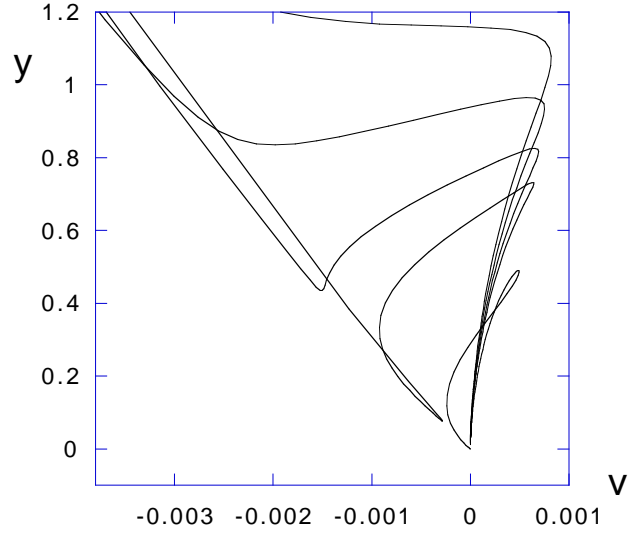}
\end{center}
\caption{Same as Fig. 3 projected on the $y$ and
$v$ plane.
\vspace{1cm}}
\end{figure}

Figures 3 and 4  show the projected solution on the $y-h_{0}$ and
$y-v$ planes, respectively. We find that $t_d = 825.7$; for
$t>t_d$ both the $y$ and $v$ variables vanish asymptotically,
while at $t<t_d$ both $y$ and $v$ are relevant. The resulting
correlation length diverges at $t_d$, similar to Fig. 2, i.e. the
transition is continuous. We have examined the transition also by
varying the initial $y_0, \, v_0$; the correlation length was
always found to diverge at $T\rightarrow T_{d}$ defining a
continuous type phase transition.

Another scenario for a 1st order transition is via changing the
initial value of $v_0$. Assuming a flow into strong renormalized
$y_r, \, v_r$ the free energy is dominated by
\begin{equation}\label{MF}
{\cal F}_r=\sum_n [-y_r\cos (\phi_n) -v_r \cos(\phi_n+\phi_{n+1})]
\end{equation}
The minimum at $\theta=\theta_n=\theta_{n+1}$ changes from
$\theta=0$ to $\cos \theta =-y_r/4v_r$ at large a negative $v_r$,
$v_r<-y_r/4$. This hints at a 1st order transition at some initial
negative $v$; this is not the decoupling transition, but rather a
transition within the coupled phase .

To emphasize the asymptotic forms we have integrated the RG
equations up to $\sqrt{y^2+v^2}=10^3$ and show in Fig. 5 the
corresponding correlation length. For these parameters, the
asymptotic v becomes negative above
$p_v=\sqrt{t|v_0|/2g}=0.5385$ (with $v_0<0$). The curve in
Fig. 5 is continuous at this $p_v$ since $x_d$ is dominated by
$y\gg v$ in the asymptotic regime. One needs to further increase
$|v_0|$ until the asymptotic $v$ and $y$ values become comparable. In
fact at $p_v=0.65$ we observe a marked change in slope for
$x_d(p_v)$ in Fig. 5. The asymptotic form $y\sim x^{\alpha}$ is
found with $\alpha$ decreasing from $1$ as $p_v$ increases,
saturating at $\alpha=1/2$ when $p_v>0.65$. Therefore, a 1st order
transition is possible within the coupled phase, associated with
the relative strength of the renormalized $y$ and $v$ variables.
The transition occurs when the initial $v_0$ is sufficiently
negative.

\begin{figure}
\begin{center}
\includegraphics[scale=0.7]{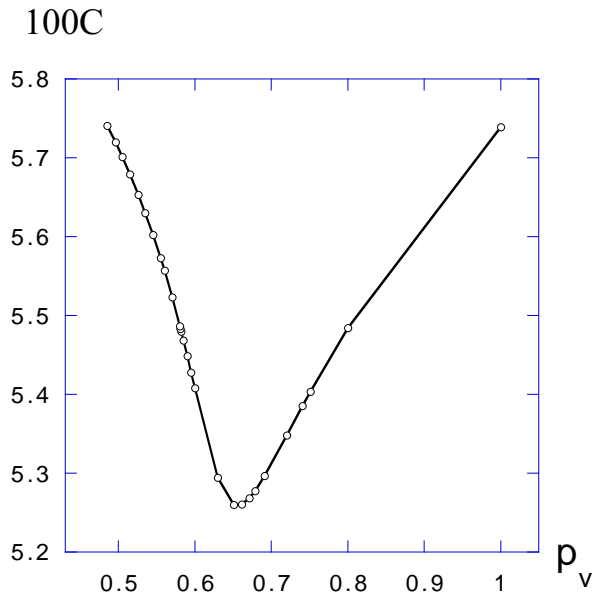}
\end{center}
\caption{The correlation length in terms of
$C=[1+2\ln (\xi_d/a)]^{-1}$ for $t=0.5$ and $p=1$
as function of $p_v=\sqrt{t|v_0|/2g}$ with $v_0<0$. The cutoff
on the RG integration of Eq. (\ref {RG1}) is $\sqrt{y^2+v^2}=10^3$.
\vspace{1cm} }
\end{figure}

\section{Plasma resonance}

We show here the relation $\omega_{pl}\sim \langle \cos \phi
\rangle$ and then apply Eq. (\ref{cosRG}) to evaluate $\langle \cos \phi
\rangle$. In presence of a weak time dependent electric field $E(t)$ in the
direction perpendicular to the layers the Josephson relation imposes a time dependent
addition $\delta \phi_n(t)$ to the Josephson phase. The kinetic energy has the form
\begin{equation}\label{Ek}
E_K=\epsilon_0\int \frac{E^2}{8\pi}d^3r=\frac{\epsilon_0
\hbar^2}{32\pi e^2d}\sum_n\int d^2r (\frac{d\delta \phi_n(t)}{dt})^2
\end{equation}
where $\epsilon_0$ is a dielectric constant. Expanding the
Josephson coupling $-(J/a^2)\sum_n\int d^2r \cos [\phi_n(\rr)+\delta
\phi_n(t)]$ yields to 2nd order $\frac{1}{2} (J/a^2)\langle \cos \phi \rangle
[\delta \phi_n(t)]^2$ (the 1st order term has $\langle \sin \phi \rangle=0$).  This form neglects the possible time dependence of the Josephson term, e.g. dynamics of pancake vortices which are assumed to be slow on the scale of $\omega_{pl}$. In particular the "phase slip" frequency was shown \cite{koshelev} to be  much smaller than $\omega_{pl}$. The plasma frequency is then
\begin{equation}\label{opl}
\omega_{pl}^2=\frac{16\pi e^2dJ}{\epsilon_0 \hbar^2a^2}\langle \cos \phi
\rangle\,.
\end{equation}

We proceed to evaluate $\langle \cos \phi \rangle$ from Eq. (\ref{cosRG}). Within 2nd order RG the contribution to the free energy \cite{h3} has the form of Eq. (\ref{dF}, i.e. $d{\cal F}=-NL^2f(x)dx$ with
\begin{equation}
f(x)=\gamma' T\{y^2(x)X_0(x)+2v^2(x)[X_0(x)+X_1(x)]\}/x^2 \,.
\end{equation}
We are mainly interested in $T>T_d$ where $v\ll y$ (see Fig. 4) so that we can use the truncated Eqs. (\ref{RG2}). We use here the solution (\ref{y}) for either $p\ll 1$ or $p\gg 1$, written as
\begin{equation}\label{yx2}
y(x)=y_0x\left(\frac{1+4gt/Z}{1+4gtx/Z}\right)^Z
\end{equation}
where $Z=t/(1+th_0^{\infty})$ and in general $Z=Z(t,J)$. The asymptotic form is $\sim x^{1-Z}$ so that $Z(t=t_d)=1$ identifies the transition temperature, e.g. in 1st order RG $Z=t$ and  (\ref{yx2}) reduces to Eq. (\ref{yx1}). Using $X_0(x)=4gtx/(1+4gtx/Z)$ and (\ref{yx2}), Eq. (\ref{cosRG}) finally yields
\begin{equation}
\langle \cos \phi \rangle =\gamma' \frac{\partial}{\partial J} \frac{J^2\tau}{2T^2}\frac{Z}{2Z-1}
\end{equation}
where $tg=T/\tau\ll 1$ is assumed, corresponding to our requirement that $T$ is well below melting.
In particular for the case $p\ll 1$, which is relevant for BSSCO \cite{shibauchi,matsuda}, we have $Z=t$ and
\begin{equation}\label{av2}
\langle \cos \phi \rangle=\gamma' \frac{J}{4gT}\frac{1}{2t-1}
\end{equation}
which can also be derived directly with (\ref{yx1}); at high temperatures $2t\gg 1$ we obtain $\langle \cos \phi \rangle=\gamma'J\tau/8T^2$. note that Eq. (\ref{av2}) is reproduced by the high temperature expansion Eq.  (\ref{av1}) up to a prefactor $4\gamma'$. While the latter expansion is in general deficient, for evaluating  $\langle \cos \phi \rangle$ it is reasonable, as discussed below Eq. (\ref{Jcor1}).

Our main result for the decoupled phase in $p\ll 1$ systems is Eq. (\ref{av2}). In comparison, the melted phase where individual pancakes are uncorrelated has a correlation length of $\sim a$, so that $\langle \cos \phi_0(0) \cos \phi_n(\rr)\rangle_c\approx e^{-r/a}$ is a plausible guess, hence from Eq. (\ref{ddJ}) we have $\langle \cos \phi \rangle\approx J/2T$; this form with a prefactor of order 1 was confirmed by simulations \cite{koshelev}. Hence in the decoupled phase $\langle \cos \phi \rangle$ is larger by a factor $\sim \tau/T$; furthermore, since $J\sim 1/B$ ($J/a^2$ is B independent) in the melted phase $\langle \cos \phi \rangle\sim 1/TB$ while in the decoupled phase it is $\langle \cos \phi \rangle\sim 1/[BT(2t-1)]$ or $\sim 1/BT^2$ not too close to decoupling. Thus the temperature dependence of $\omega_{pl}$ can  distinguish between decoupled and liquid phases.

\section{The SCHA method}

In this section we derive the decoupling transition within the
variational SCHA method, reproducing the results of
ref.\cite{daemen}. In particular, this method results in a 1st
order transition at large $p$. We compare the method to that of
2nd order RG and show where the deficiency of SCHA originates.

The SCHA proceeds by searching for the optimal Gaussian
Hamiltonian of the form
\begin{equation}
{\cal H}_0=\frac{1}{2}\int_{\qq,k}\sum_{i,j}
G_s^{-1}(q,k)\phi(\qq,k)\phi^{*}(\qq,k)
\end{equation}
so that $G_s(q,k)$ is determined by minimization of the
 variational free energy
\[F_{var}=F_{0}+<H-H_{0}>_{0}\]
where the averaging as well as $F_0$ correspond to ${\cal H}_0$.
$F_{var}$ is then
\begin{equation}\label{Fvar}
{\cal F}_{var}=\frac{1}{2}\int_{q,k}\{-\ln
G_s(q,k)+[G_s^{-1}(q,k)-G_b^{-1}(q,k)]G_s(q,k)\}
-\frac{J}{a^2}e^{-\frac{1}{2}\int_{q,k}G_s(q,k)}
\end{equation}
where the $\ln G_s$ term corresponds to $F_0$ and we have used
$<\phi^*(\qq,k)\phi(\qq,k)>_{0}=TG_s(q,k)$. The last term in (\ref
{Fvar}) is the Josephson term with,
\begin{equation}
<\cos
\phi(\rr)>_{0}=e^{-\frac{1}{2}T\int_{q,k}G_s(q,k)}=exp(-s_v)\, .
\label{sv}
\end{equation}
This defines a parameter $s_v$; the renormalized Josephson
coupling is then $Je^{-s_v}$.

 Minimizing Eq. (\ref{Fvar}) yields
\begin{equation}
G_s^{-1}=G_b^{-1}+\frac{J}{a^2d} \exp(-s_v)  \label{GMF}
\end{equation}
Using the form  (\ref{Gb1}) and the variable $x$ with
$d^2q/(2\pi)^2=-dx/(ax)^2$, Eqs. (\ref{sv},\ref{GMF}) reduce to a
self consistent equation for $s_v$
\begin{equation}\label{sv1}
s_v=d\int \frac{dk}{2\pi} \int_1^{\infty} dx
\frac{1}{\frac{x+1/4g}{2t\sin^2(kd/2)}
+\frac{2J}{T}x^2e^{-s_v}}\,.
\end{equation}
This equation has exactly the same structure as that of Eq.
(\ref{s}) within 2nd order RG if the asymptotic form of RG
variable $h_0(x)\rightarrow xy_0e^{-s}$ is used. However the
detailed $h_0(x)$ behavior is significant and can affect the
critical properties; in fact, $y$ and $h_0(x)$ are non-monotonic.

We can perform the $k$ integration in Eq. (\ref{sv1}), neglecting
the $f$ type function as in Eq. (\ref{X02}) (i.e. replacing
$0.84<f<1$ by $f=1$) leading to
\begin{equation}\label{sv2}
s_v=\int_1^{\infty} dx\frac{4gt}{1+4gx
+(4g)^2p^2x^2e^{-s_v}}= \left\{\begin{array}{ll}
\frac{8gt}{\sqrt{D}} \arctan
\frac{2/x+4g}{\sqrt{D}} |_{\infty}^{1}&
\mbox{if $D > 0$}\\
\frac{4gt}{\sqrt{-D}} ln\frac{2/x+4g-\sqrt{-D}}
{2/x+4g+\sqrt{-D}} |_{\infty}^{1} & \mbox{if $D < 0$}
\end{array}
\right.
\end{equation}
where $D=16g^{2}(4p^{2} \exp(-s_v)-1)$ and $p$ is defined
in Eq. (\ref{p}).

If the transition is continuous then $s_v$ diverges near $T_d$ so
that the effective Josephson coupling $\sim e^{-s_v}\rightarrow 0$;
Eq. (\ref{sv2}) then yields $t_d=1$. The RG result shows instead a
weak p dependence even at small $p$ as in Eq. (\ref{tds}) and Fig.
1.

At a 1st order transition $s_v$ is finite; anticipating a large $p$,
$4p^{2} \exp(-s_v) \gg 1$ but $D\approx (8gp)^2 e^{-s_v}\ll
1$,  Eq.(\ref{sv2}) can be written as:
\begin{equation}
s_v e^{-s_v/2}=\frac{\pi t}{2p}\,.
\end{equation}
The product $s_ve^{-s_v/2}$ is bounded by $2/e$ at $s_v=2$, hence as
temperature approaches $T_d$ from below $s_v$ increases up to $s_v=2$
but then jumps to $s_v=\infty$ in the decoupled phase, i.e. a 1st
order transition. The critical temperature is then
\begin{equation}\label{Td2}
t_d=\frac{4}{\pi e}p  \qquad  1 \ll p \ll 1/g
\end{equation}
This result is similar to that from RG, Eq. (\ref{tdb}), except
that the slope is somewhat different. The significant difference
is that RG yields a continuous transition even at large $p$.

At even larger p, where $p\gg 1/g$, $D\gg 1$, Eq.
(\ref{sv2}) yields
\begin{equation}
s_v e^{-s_v}=\frac{ t}{4g p^2}\,.
\end{equation}
As above, $s_v e^{-s_v}$ is bounded by $1/e$ at $s_v=1$, hence a 1st
order transition at
\begin{equation}\label{Td3}
t_d=\frac{4}{e}gp^2  \qquad   p \gg 1/g
\end{equation}

The results $t_d=1$ for weak $p$ and Eq. (\ref{Td3}) for strong
$p$ are the results given in Ref. \cite{daemen}. The intermediate
range Eq. (\ref{Td2}) is not mentioned there, though the plotted
decoupling fields $B_D(T)$ in their Fig. 1 are consistent with
$B_D(T)\sim 1/T^2$ as from Eq. (\ref{Td2}). Furthermore, Eq.
(\ref{Td3}) yields $T_d=(4gp)^2\tau/4e\gg \tau$ which is
incompatible with the requirement that $T_d$ is well below $T_m$.

It is interesting to note that the early estimate of Glazman and
Koshelev \cite{glazman2} of the decoupling temperature gives a
result similar to (\ref{Td2}) or (\ref{tdb}). Within this estimate
the $\cos\phi$ term in Eq. (\ref{H}) is expanded and the condition
of large fluctuations $\langle \phi^2\rangle \approx 1$ with
$\langle \phi^2\rangle=T\int_{q,k}[G_b^{-1}(q,k)+J/a^2]^{-1}$
yields $T_d$. (This condition indicates decoupling, though by
itself does not prove a phase transition.) The result is then the
same as Eqs. (\ref{sv},\ref{GMF}) with $s_v\approx 1$, therefore
it yields indeed a result close to that of Eq. (\ref{Td2}).

Our main result in this section is to show the formal similarity
between SCHA and 2nd order RG as well as an important difference,
i.e. the mass term which is generated by RG is scale dependent.
Both methods show significant enhancement of $T_d$ with increasing
Josephson coupling, however the transition remains continuous in
the RG solution.

\section{Conclusions}

In recent experiments on BSCCO \cite{khaykovich,fuchs} the phase
diagram has shown a number of low temperature phases. Most of
these transitions are disorder driven by either bulk pinning or by
surface barriers. In particular the possibility that the second
peak transition is a disorder driven decoupling has been suggested
\cite{h1,h2}. The significant reduction of the Josephson plasma
resonance at the second peak supports a decoupling scenario
\cite{shibauchi,matsuda}, however a conclusive signature for
decoupling has not been shown so far.

The signature of decoupling is that translational order is
maintained, though with softer tilt modulous, while
superconducting order is lost, i.e. $Q=0$ (Eq. \ref{Q}) and the critical current in the
$c$ direction vanish at $T>T_d$. An additional signature is
the power law decay of the Josephson correlation at $T>T_d$.
We find that the decoupling transition temperature is
$T_d\sim1/B$ at low fileds ($T_d^0$ of Eq. \ref {Td00}) while it changes
to $T_d=0.95T_d^0p\sim 1/\sqrt{B}$ at higher fields ($a\lambda_J\lesssim \lambda^2$)
with significantly enhanced temperatures $T_d\gg T_d^0$.

We have shown that RG can be used to evaluate $\langle \cos \phi \rangle$ and hence the Josephson plasma frequency. In particular for weak coupling $J$, as in BSCCO, we find that $\omega_{pl}\sim 1/[BT(2T/T_d^0-1)]$, in contrast to a $\sim 1/BT$ behavior in the melted phase. This temperature dependence can serve to identify a decoupled phase.

In the present work we assume that V-I defects are not generated.
Hence superconductivity is lost only in the $c$ direction (i.e.
$I_c=0$) while 2-dimensional superconductivity is maintained
parallel to the layers. Our neglect of V-I defects is in fact not
justified, since they are generated at a lower temperature in the
$J=0$ system \cite{dodgson,ledou}, i.e. at $T_d^0/8$. The true
transition is a 3-dimensional one in which both decoupling and the
defect transition coalesce, similar to the $B=0$ scenario
\cite{h3}. The actual transition temperature $T_c$ is between
$T_d$ and the defect transition. It can be estimated by the
temperature at which the correlation lengths of the defects
$\xi_{def}$ and $\xi_d$ become comparable. Thus e.g., if
$\xi_{def}>\xi_d$, the Josephson coupling is renormalized to
strong coupling before the $\cos$ term feels the V-I defects. Since
$\xi_{def}\approx a\exp (E_c/T)$ (where $E_c\approx 0.2\tau$ is
the pancake vortex core energy; if local lattice relaxation is
included \cite{olive} then $E_c\approx 0.04\tau$) $\xi_{def}$ is
exponentially large, and $T_c$ is close to $T_d$ unless $J$ is
extremely small.

We expect for systems like BSCCO or YBCO that decoupling affects
mostly superconductivity in the $c$ direction. The current-voltage relation
parallel to the layers is expected then to be nonlinear \cite{h1},
except at very low currents where the few V-I defects would
eventually lead to a linear Ohmic behavior. Similarly, the power
law for the Josephson correlation would eventually, beyond the V-I
spacing $\xi_{def}$  decay exponentially.

In conclusion we have studied the meaning and critical properties
of the decoupling transition. On the theory side, in our view this
is one of the few transitions of vortex matter which is fully
understood. It remains to be seen if experiment can also provide
clear realizations for this type of transition.

\begin{acknowledgments}
This research was supported by THE ISRAEL SCIENCE FOUNDATION
founded by the Israel Academy of Sciences and Humanities. We thank A. Aharony, G. Zar\'{a}nd and S. Teitel for useful and valuable comments.
\end{acknowledgments}

\appendix

\section{Expansions for $T_d$}

We present in this appendix an analytic expansion for the
decoupling temperature $T_d$ within the reduced set of Eq.
(\ref{yh0}) with $f(h_0,x,t)=1$. The results show a significant
enhancement when the parameter $p$ in Eq. (\ref{p}) is large.

We consider $T>T_d$ where $y$ flows to zero and $h_0$ reaches a
finite asymptotic value $h_{0}^{\infty}$, since the integration of
$dh_0$ in Eq. (\ref{yh0}) is convergent. We assume that the
integration of the $y$ equation is dominated by $h^{0}_{\infty}$
and will examine below the validity of this assumption.

Integrating $y(x)$ in Eq.(\ref{yh0}) with $h_0\rightarrow
h_{0}^{\infty}$ yields
\begin{equation}
y=y_{0}x\left(\frac{1+4g(1+t
h_{0}^{\infty})}{1+4gx(1+t h_{0}^{\infty})}\right)^{
\frac{t}{ 1+t h_{0}^{\infty}} } \,.\label{y}
\end{equation}
We now substitute this $y(x)$ solution into the $h_0$ equation
(\ref{yh0}) and solve for $h_0$,
\begin{eqnarray}\label{h00}
h_{0}(x)=\frac{p^{4}Z^{2}}{
 t^{4} (Z-1)(2 Z-1)}&&\left(
1-\frac{Z(2Z-1)}{(1+4gt x/Z)^{2(Z-1)}}+
\frac{4Z(Z-1)}{(1+4gt x/Z)^{2Z-1}}\right.\nonumber\\
&&\left. -\frac{(t/p-1)(2t/p-1)}{(1+4gt x)^{2t/p}}
\right)
\end{eqnarray}
where  $h_{0}^{\infty}$ is also included in $Z=1/(1/t+
h_{0}^{\infty})$ and the important parameter $p$ is defined in Eq.
(\ref{p}). Since $y\sim x^{1-Z}$ at large $x$, $Z=1$ defines the transition point.

We can now examine the consistency of replacing $h_0(x)$ by
$h_{0}^{\infty}$ in the $y$ equation. The condition that $h_0(x)$
approaches $h_{0}^{\infty}$ relatively fast is that the $x$
dependent terms in Eq. (\ref{h00}) are small at $x>1/4gt$,
i.e. $Z$ is small (but $Z>1$), hence $p$ and $h^{0}_{\infty}$ are
large. It seems plausible then that our approximation is valid at
$p\gg 1$. Alternatively, if $h_0(x)\ll 1$ then it has anyway a
weak effect on the RG, i.e. the present derivation is valid at
$p\ll 1$.

Eq. (\ref{h00}) shows that $h_0(x)$ converges to $h^{0}_{\infty}$
if $Z>1$. We now substitute $x\rightarrow \infty$ in (\ref{h00})
and obtain a self consistent equation for $h^{0}_{\infty}$, which
for the variable $Z$ becomes a cubic equation
\begin{equation}
 Z^{3}-\frac{2+3/t}{(p/t)^{4}+2/t}Z^{2}+\frac{3+1/t}{(p/t)^{4}+2/t}Z-
\frac{1}{(p/t)^{4}+2/t}=Z^{3}+a_{2}Z^{2}+a_{1}Z+a_{0}=0
\label{cubic}
\end{equation}
This cubic equation has  solutions $Z>1$  only if the condition
$D(t)<0$ is satisfied, where
\begin{equation}
D(t)=\frac{1}{27}a_{1}^{3}-
\frac{1}{6}a_{0}a_{1}a_{2}+\frac{1}{27}a_{0}a_{2}^{3}-\frac{1}{4\cdot
27} a_{1}^{2}a_{2}^{2}+\frac{1}{4}a_{0}^{2}\,.
\end{equation}
Therefore, $D(t)=0$ correspond to $Z=1$ and defines the
temperature of the phase transition $T_{d}$ as given in Eqs.
(\ref{tds},\ref{tdb}).

\end{document}